\documentstyle[12pt]{article} 

\def\doit#1#2{\ifcase#1\or#2\fi} 

\expandafter\ifx\csname amsppt.sty\endcsname\endinput
  \expandafter\def\csname amsppt.sty\endcsname{2.2 (2001/08/07)}\fi

\catcode`@=11 
\catcode`@=12 

\let\du=\d                      

\def\a{\alpha} \def\b{\beta}  \def\d{\delta}
\def\e{\epsilon}  \def\g{\gamma}
   \def\k{\kappa}
\def\l{\lambda} \def\m{\mu} \def\n{\nu} 
  \def\r{\rho} \def\s{\sigma}

\def\pmb#1{\setbox0=\hbox{${#1}$}%
   \kern-.025em\copy0\kern-\wd0
   \kern-.035em\copy0\kern-\wd0
   \kern.05em\copy0\kern-\wd0
   \kern-.035em\copy0\kern-\wd0
   \kern-.025em\box0 }

\def\bo{{\raise-.46ex\hbox{\large$\Box$}}} 

\def\pr{\prod}                            

\def\TH{{\raise.2ex\hbox{$\displaystyle \bigodot$}\mskip-4.7mu %
\llap H \;}}
\def\face{{\raise.2ex\hbox{$\displaystyle \bigodot$}\mskip-2.2mu %
\llap {$\ddot
        \smile$}}}                           

\def\sp#1{{}^{#1}}                 

\def\Tilde#1{{\widetilde{#1}}\hskip 0.015in}     
\def\Hat#1{\widehat{#1}}                        
\def\Bar#1{\overline{#1}}                       
\def\leftrightarrowfill{$\mathsurround=0pt \mathord\leftarrow 
 \mkern-6mu
        \cleaders\hbox{$\mkern-2mu \mathord- \mkern-2mu$}\hfill
        \mkern-6mu \mathord\rightarrow$}
\def\dvec#1{\vbox{\ialign{##\crcr
        \leftrightarrowfill\crcr\noalign{\kern-1pt\nointerlineskip}
        $\hfil\displaystyle{#1}\hfil$\crcr}}}           
\def\dt#1{{\buildrel {\hbox{\LARGE .}} \over {#1}}}

\def\frac#1#2{{\textstyle{#1\over\vphantom2\smash{\raise.20ex
        \hbox{$\scriptstyle{#2}$}}}}}   
\def\sfrac#1#2{{\vphantom1\smash{\lower.5ex\hbox{\small$#1$}}\over
        \vphantom1\smash{\raise.4ex\hbox{\small$#2$}}}}
\def\bfrac#1#2{{\vphantom1\smash{\lower.5ex\hbox{$#1$}}\over
        \vphantom1\smash{\raise.3ex\hbox{$#2$}}}}       
\def\afrac#1#2{{\vphantom1\smash{\lower.5ex\hbox{$#1$}}\over#2}} 
\def\on#1#2{\mathop{\null#2}\limits^{#1}}       

\newskip\humongous \humongous=0pt plus 1000pt minus 1000pt
\def\caja{\mathsurround=0pt}

\newif\ifdtup
\def\panorama{\global\dtuptrue \openup2\jot \caja
        \everycr{\noalign{\ifdtup \global\dtupfalse
        \vskip-\lineskiplimit \vskip\normallineskiplimit
        \else \penalty\interdisplaylinepenalty \fi}}}
\def\li#1{\panorama \tabskip=\humongous      
        \halign to\displaywidth{\hfil$\displaystyle{##}$
        \tabskip=0pt&$\displaystyle{{}##}$\hfil
        \tabskip=\humongous&\llap{$##$}\tabskip=0pt
        \crcr#1\crcr}}

\doit0{
\def\ref#1{$\sp{#1)}$}
}

\parskip=\medskipamount          
\def\baselinestretch{1.2}       
\thicklines                         

\def\endtitle{\end{quotation}\newpage}  

\def\sect#1{\bigskip\medskip \goodbreak \noindent{\bf {#1}} %
\nobreak \medskip}
\def\refs{\sect{References} \footnotesize \frenchspacing \parskip=0pt}
\def\Item{\par\hang\textindent}

\def\[{\lfloor{\hskip 0.35pt}\!\!\!\lceil}
\def\]{\rfloor{\hskip 0.35pt}\!\!\!\rceil}

\def\nablasl{{{\nabla{\hskip -7.7pt}\raise 1.5pt \hbox{/} \,}}}

\def\Lag{{\cal L}}
\def\du#1#2{_{#1}{}^{#2}}

\def\calD{{\cal D}}
\def\calG{{\cal G}}

\def\rma{{\rm a}} \def\rmb{{\rm b}} \def\rmc{{\rm c}} 
\def\rmd{{\rm d}} 
\def\rme{{\rm e}} \def\rmf{{\rm f}} \def\rmg{{\rm g}} 
\def\rmh{{\rm h}}

\def\plpl{{+\!\!\!\!\!{\hskip 0.009in}%
{\raise-1.0pt\hbox{$_+$}}  {\hskip 0.0008in}}} 
\def\mimi{{-\!\!\!\!\!{\hskip 0.009in}%
{\raise-1.0pt\hbox{$_-$}}  {\hskip 0.0008in}}}

\def\order#1#2{{\cal O}({#1}^{#2})}

\def\pl#1#2#3{Phys.~Lett.~{\bf {#1}B} (19{#2}) #3}
\def\np#1#2#3{Nucl.~Phys.~{\bf B{#1}} (19{#2}) #3}
\def\prl#1#2#3{Phys.~Rev.~Lett.~{\bf #1} (19{#2}) #3}
\def\pr#1#2#3{Phys.~Rev.~{\bf D{#1}} (19{#2}) #3}

\def\ap#1#2#3{Ann.~of Phys.~{\bf {#1}} (19{#2}) #3} 
\def\prep#1#2#3{Phys.~Rep.~{\bf {#1}} (19{#2}) #3}
\def\jhep#1#2#3{JHEP {\bf {#1}} (19{#2}) #3}
\def\ptp#1#2#3{Prog.~Theor.~Phys.~{\bf {#1}} (19{#2}) #3}

\def\nc#1#2#3{Nuovo Cim.~{\bf {#1}} (19{#2}) #3}
\def\ibid#1#2#3{{\it ibid.}~{\bf {#1}} (19{#2}) #3}

\def\mpl#1#2#3{Mod.~Phys.~Lett.~{\bf A{#1}} (19{#2}) #3}

\def\hepth#1{{hep-th/{#1}}} 
 
\def\hepph#1{{hep-ph/{#1}}} 
\def\texttts#1{\small\texttt{#1}} 
\def\arXive#1{arXiv:{#1}{$\,$}[hep-th]}

\def\pln#1#2#3{Phys.~Lett.~{\bf {#1}B} (20{#2}) #3} 
\def\npn#1#2#3{Nucl.~Phys.~{\bf B{#1}} (20{#2}) #3}
\def\prln#1#2#3{Phys.~Rev.~Lett.~{\bf #1} (20{#2}) #3}
\def\prn#1#2#3{Phys.~Rev.~{\bf D{#1}} (20{#2}) #3}

\def\prepn#1#2#3{Phys.~Rep.~{\bf {#1}C} (20{#2}) #3}
\def\jhepn#1#2#3{JHEP {\bf {#1}} (20{#2}) #3}
\def\ptpn#1#2#3{Prog.~Theor.~Phys.~{\bf {#1}} (20{#2}) #3}
\def\ijmpn#1#2#3{Int.~Jour.~Mod.~Phys.~{\bf A{#1}} (20{#2}) #3}

\def\mpln#1#2#3{Mod.~Phys.~Lett.~{\bf A{#1}} (20{#2}) #3}

\def\<<{<\!\!<} \def\>>{>\!\!>} 
\def\Check#1{{\raise-1.0pt\hbox{\LARGE\v{}}{\hskip -10pt}{#1}}}

\def\eqques{{~\,={\hskip -11.5pt}\raise -1.8pt\hbox{\large ?}
{\hskip 4.5pt}}{}}
\def\fracm#1#2{\,\hbox{\large{${\frac{{#1}}{{#2}}}$}}\,}
\def\fracmm#1#2{\,{{#1}\over{#2}}\,}

\def\frac#1#2{{\textstyle{#1\over\vphantom2\smash{\raise -.20ex
        \hbox{$\scriptstyle{#2}$}}}}}   
\def\sqrttwo{{\sqrt2}}
\def\scst{\scriptstyle}

\def\.{.$\,$}
\def\-{{\hskip 1.5pt}\hbox{-}}

\def\footnotes#1{{\hskip 1pt}\footnotemark$^)$\footnotetext%
{\hsize=6.5in $^)$~{#1}}} 

\def\low#1{\hskip0.01in{\raise -3pt\hbox{${\hskip 1.0pt}\!_{#1}$}}}
\def\low#1{\hskip0.01in{\raise -3pt\hbox{$\!\!\!_{#1}$}}}
\def\ip{{=\!\!\! \mid}}

\begin{document}

\font\tenmib=cmmib10
\font\sevenmib=cmmib10 at 7pt 
\font\fivemib=cmmib10 at 5pt  
\font\tenbsy=cmbsy10
\font\sevenbsy=cmbsy10 at 7pt 
\font\fivebsy=cmbsy10 at 5pt  
\def\BMfont{\textfont0\tenbf \scriptfont0\sevenbf
                              \scriptscriptfont0\fivebf
            \textfont1\tenmib \scriptfont1\sevenmib
                               \scriptscriptfont1\fivemib
            \textfont2\tenbsy \scriptfont2\sevenbsy
                               \scriptscriptfont2\fivebsy}
\def\rlx{\relax\leavevmode}                  
\def\BM#1{\rlx\ifmmode\mathchoice
                      {\hbox{$\BMfont#1$}}
                      {\hbox{$\BMfont#1$}}
                      {\hbox{$\scriptstyle\BMfont#1$}}
                      {\hbox{$\scriptscriptstyle\BMfont#1$}}
                 \else{$\BMfont#1$}\fi}

\font\tenmib=cmmib10
\font\sevenmib=cmmib10 at 7pt 
\font\fivemib=cmmib10 at 5pt  
\font\tenbsy=cmbsy10
\font\sevenbsy=cmbsy10 at 7pt 
\font\fivebsy=cmbsy10 at 5pt  
\def\BMfont{\textfont0\tenbf \scriptfont0\sevenbf
                              \scriptscriptfont0\fivebf
            \textfont1\tenmib \scriptfont1\sevenmib
                               \scriptscriptfont1\fivemib
            \textfont2\tenbsy \scriptfont2\sevenbsy
                               \scriptscriptfont2\fivebsy}
\def\BM#1{\rlx\ifmmode\mathchoice
                      {\hbox{$\BMfont#1$}}
                      {\hbox{$\BMfont#1$}}
                      {\hbox{$\scriptstyle\BMfont#1$}}
                      {\hbox{$\scriptscriptstyle\BMfont#1$}}
                 \else{$\BMfont#1$}\fi}

\def\inbar{\vrule height1.5ex width.4pt depth0pt}
\def\sinbar{\vrule height1ex width.35pt depth0pt}
\def\ssinbar{\vrule height.7ex width.3pt depth0pt}
\font\cmss=cmss10
\font\cmsss=cmss10 at 7pt
\def\ZZ{{}Z {\hskip -6.7pt} Z{}} 
\def\Ik{\rlx{\rm I\kern-.18em k}}  
\def\IC{\rlx\leavevmode
             \ifmmode\mathchoice
                    {\hbox{\kern.33em\inbar\kern-.3em{\rm C}}}
                    {\hbox{\kern.33em\inbar\kern-.3em{\rm C}}}
                    {\hbox{\kern.28em\sinbar\kern-.25em{\rm C}}}
                    {\hbox{\kern.25em\ssinbar\kern-.22em{\rm C}}}
             \else{\hbox{\kern.3em\inbar\kern-.3em{\rm C}}}\fi}
\def\IP{\rlx{\rm I\kern-.18em P}}
\def\IR{\rlx{\rm I\kern-.18em R}}
\def\IN{\rlx{\rm I\kern-.20em N}}
\def\Ione{\rlx{\rm 1\kern-2.7pt l}}
\def\bbbzz{{\Bbb Z}}

%
\def\unredoffs{} \def\redoffs{\voffset=-.31truein\hoffset=-.59truein}
\def\speclscape{\special{ps: landscape}}

\newbox\leftpage \newdimen\fullhsize \newdimen\hstitle\newdimen\hsbody
\tolerance=1000\hfuzz=2pt\def\fontflag{cm}
\catcode`\@=11 
\hsbody=\hsize \hstitle=\hsize 

\def\nolabels{\def\wrlabeL##1{}\def\eqlabeL##1{}\def\reflabeL##1{}}
\def\writelabels{\def\wrlabeL##1{\leavevmode\vadjust{\rlap{\smash%
{\line{{\escapechar=` \hfill\rlap{\sevenrm\hskip.03in\string##1}}}}}}}%
\def\eqlabeL##1{{\escapechar-1\rlap{\sevenrm\hskip.05in\string##1}}}%
\def\reflabeL##1{\noexpand\llap{\noexpand\sevenrm\string\string%
\string##1}}}
\nolabels
%
\global\newcount\secno \global\secno=0
\global\newcount\meqno \global\meqno=1
\def\newsec#1{\global\advance\secno by1\message{(\the\secno. #1)}
\global\subsecno=0\eqnres@t\noindent{\bf\the\secno. #1}
\writetoca{{\secsym} {#1}}\par\nobreak\medskip\nobreak}
\def\eqnres@t{\xdef\secsym{\the\secno.}\global\meqno=1
\bigbreak\bigskip}
\def\sequentialequations{\def\eqnres@t{\bigbreak}}\xdef\secsym{}
\global\newcount\subsecno \global\subsecno=0
\def\subsec#1{\global\advance\subsecno by1%
\message{(\secsym\the\subsecno.%
 #1)}
\ifnum\lastpenalty>9000\else\bigbreak\fi
\noindent{\it\secsym\the\subsecno. #1}\writetoca{\string\quad
{\secsym\the\subsecno.} {#1}}\par\nobreak\medskip\nobreak}
\def\appendix#1#2{\global\meqno=1\global\subsecno=0%
\xdef\secsym{\hbox{#1.}}
\bigbreak\bigskip\noindent{\bf Appendix #1. #2}\message{(#1. #2)}
\writetoca{Appendix {#1.} {#2}}\par\nobreak\medskip\nobreak}
\def\eqnn#1{\xdef #1{(\secsym\the\meqno)}\writedef{#1\leftbracket#1}%
\global\advance\meqno by1\wrlabeL#1}
\def\eqna#1{\xdef #1##1{\hbox{$(\secsym\the\meqno##1)$}}
\writedef{#1\numbersign1\leftbracket#1{\numbersign1}}%
\global\advance\meqno by1\wrlabeL{#1$\{\}$}}
\def\eqn#1#2{\xdef #1{(\secsym\the\meqno)}\writedef{#1\leftbracket#1}%
\global\advance\meqno by1$$#2\eqno#1\eqlabeL#1$$}
%
\newskip\footskip\footskip8pt plus 1pt minus 1pt 
\font\smallcmr=cmr5 
\def\footnotefont{\smallcmr}
\def\f@t#1{\footnotefont #1\@foot}
\def\f@@t{\baselineskip\footskip\bgroup\footnotefont\aftergroup%
\@foot\let\next}
\setbox\strutbox=\hbox{\vrule height9.5pt depth4.5pt width0pt} %
\global\newcount\ftno \global\ftno=0
\def\foot{\global\advance\ftno by1\footnote{$^{\the\ftno}$}}
%
\newwrite\ftfile
\def\footend{\def\foot{\global\advance\ftno by1\chardef\wfile=\ftfile
$^{\the\ftno}$\ifnum\ftno=1\immediate\openout\ftfile=foots.tmp\fi%
\immediate\write\ftfile{\noexpand\smallskip%
\noexpand\item{f\the\ftno:\ }\pctsign}\findarg}%
\def\footatend{\vfill\eject\immediate\closeout\ftfile{\parindent=20pt
\centerline{\bf Footnotes}\nobreak\bigskip\input foots.tmp }}}
\def\footatend{}
\global\newcount\refno \global\refno=1
\newwrite\rfile
%
\def\ref{[\the\refno]\nref}%
\def\nref#1{\xdef#1{[\the\refno]}\writedef{#1\leftbracket#1}%
\ifnum\refno=1\immediate\openout\rfile=refs.tmp\fi%
\global\advance\refno by1\chardef\wfile=\rfile\immediate%
\write\rfile{\noexpand\Item{#1}\reflabeL{#1\hskip.31in}\pctsign}%
\findarg\hskip10.0pt}%
\def\findarg#1#{\begingroup\obeylines\newlinechar=`\^^M\pass@rg}
{\obeylines\gdef\pass@rg#1{\writ@line\relax #1^^M\hbox{}^^M}%
\gdef\writ@line#1^^M{\expandafter\toks0\expandafter{\striprel@x #1}%
\edef\next{\the\toks0}\ifx\next\em@rk\let\next=\endgroup%
\else\ifx\next\empty%
\else\immediate\write\wfile{\the\toks0}%
\fi\let\next=\writ@line\fi\next\relax}}
\def\striprel@x#1{} \def\em@rk{\hbox{}}
\def\lref{\begingroup\obeylines\lr@f}
\def\lr@f#1#2{\gdef#1{\ref#1{#2}}\endgroup\unskip}
\def\semi{;\hfil\break}
\def\addref#1{\immediate\write\rfile{\noexpand\item{}#1}} 
%
\def\listrefs{\footatend\vfill\supereject\immediate\closeout%
\rfile\writestoppt
\baselineskip=14pt\centerline{{\bf References}}%
\bigskip{\frenchspacing%
\parindent=20pt\escapechar=` \input refs.tmp%
\vfill\eject}\nonfrenchspacing}
%
\def\listrefsr{\immediate\closeout\rfile\writestoppt
\baselineskip=14pt\centerline{{\bf References}}%
\bigskip{\frenchspacing%
\parindent=20pt\escapechar=` \input refs.tmp\vfill\eject}%
\nonfrenchspacing}
\def\listrefsrsmall{\immediate\closeout\rfile\writestoppt
\baselineskip=11pt\centerline{{\bf References}} 
\font\smallerfonts=cmr9 \font\it=cmti9 \font\bf=cmbx9%
\bigskip{\smallerfonts{%
\parindent=15pt\escapechar=` \input refs.tmp\vfill\eject}}}
\def\listrefsrmed{\immediate\closeout\rfile\writestoppt
\baselineskip=12.5pt\centerline{{\bf References}}
\font\smallerfonts=cmr10 \font\it=cmti10 \font\bf=cmbx10%
\bigskip{\smallerfonts{%
\parindent=18pt\escapechar=` \input refs.tmp\vfill\eject}}}
\def\startrefs#1{\immediate\openout\rfile=refs.tmp\refno=#1}
\def\xref{\expandafter\xr@f}\def\xr@f[#1]{#1}
\def\refs#1{\count255=1[\r@fs #1{\hbox{}}]}
\def\r@fs#1{\ifx\und@fined#1\message{reflabel %
\string#1 is undefined.}%
\nref#1{need to supply reference \string#1.}\fi%
\vphantom{\hphantom{#1}}\edef\next{#1}\ifx\next\em@rk\def\next{}%
\else\ifx\next#1\ifodd\count255\relax\xref#1\count255=0\fi%
\else#1\count255=1\fi\let\next=\r@fs\fi\next}
\def\figures{\centerline{{\bf Figure Captions}}%
\medskip\parindent=40pt%
\def\fig##1##2{\medskip\item{Fig.~##1.  }##2}}
%

\newwrite\ffile\global\newcount\figno \global\figno=1
\doit0{
\def\fig{fig.~\the\figno\nfig}
\def\nfig#1{\xdef#1{fig.~\the\figno}%
\writedef{#1\leftbracket fig.\noexpand~\the\figno}%
\ifnum\figno=1\immediate\openout\ffile=figs.tmp%
\fi\chardef\wfile=\ffile%
\immediate\write\ffile{\noexpand\medskip\noexpand%
\item{Fig.\ \the\figno. }
\reflabeL{#1\hskip.55in}\pctsign}\global\advance\figno by1\findarg}
\def\listfigs{\vfill\eject\immediate\closeout\ffile{\parindent40pt
\baselineskip14pt\centerline{{\bf Figure Captions}}\nobreak\medskip
\escapechar=` \input figs.tmp\vfill\eject}}
\def\xfig{\expandafter\xf@g}\def\xf@g fig.\penalty\@M\ {}
\def\figs#1{figs.~\f@gs #1{\hbox{}}}
\def\f@gs#1{\edef\next{#1}\ifx\next\em@rk\def\next{}\else
\ifx\next#1\xfig #1\else#1\fi\let\next=\f@gs\fi\next}
}

\newwrite\lfile
{\escapechar-1\xdef\pctsign{\string\%}\xdef\leftbracket{\string\{}
\xdef\rightbracket{\string\}}\xdef\numbersign{\string\#}}
\def\writedefs{\immediate\openout\lfile=labeldefs.tmp %
\def\writedef##1{%
\immediate\write\lfile{\string\def\string##1\rightbracket}}}
\def\writestop{\def\writestoppt%
{\immediate\write\lfile{\string\pageno%
\the\pageno\string\startrefs\leftbracket\the\refno\rightbracket%
\string\def\string\secsym\leftbracket\secsym\rightbracket%
\string\secno\the\secno\string\meqno\the\meqno}%
\immediate\closeout\lfile}}
\def\writestoppt{}\def\writedef#1{}
\def\seclab#1{\xdef #1{\the\secno}\writedef{#1\leftbracket#1}%
\wrlabeL{#1=#1}}
\def\subseclab#1{\xdef #1{\secsym\the\subsecno}%
\writedef{#1\leftbracket#1}\wrlabeL{#1=#1}}
\newwrite\tfile \def\writetoca#1{}
\def\leaderfill{\leaders\hbox to 1em{\hss.\hss}\hfill}
\def\writetoc{\immediate\openout\tfile=toc.tmp
   \def\writetoca##1{{\edef\next{\write\tfile{\noindent ##1
   \string\leaderfill {\noexpand\number\pageno} \par}}\next}}}
\def\listtoc{\centerline{\bf Contents}\nobreak%
 \medskip{\baselineskip=12pt
 \parskip=0pt\catcode`\@=11 \input toc.tex \catcode`\@=12 %
 \bigbreak\bigskip}} 
\catcode`\@=12 
%

\countdef\pageno=0 \pageno=1
\newtoks\headline \headline={\hfil} 
\newtoks\footline 
 \footline={\bigskip\hss\tenrm\folio\hss}
\def\folio{\ifnum\pageno<0 \romannumeral-\pageno \else\number\pageno 
 \fi} 

\def\nopagenumbers{\footline={\hfil}} 
\def\advancepageno{\ifnum\pageno<0 \global\advance\pageno by -1 
 \else\global\advance\pageno by 1 \fi} 
\newif\ifraggedbottom

\def\raggedbottom{\topskip10pt plus60pt \raggedbottomtrue}
\def\normalbottom{\topskip10pt \raggedbottomfalse} 

\def\on#1#2{{\buildrel{\mkern2.5mu#1\mkern-2.5mu}\over{#2}}}
\def\dt#1{\on{\hbox{\bf .}}{#1}}                
\def\Dot#1{\dt{#1}}

\def\eqdot{{\hskip4pt}{\buildrel{\hbox{\LARGE .}} \over =}\,\,{}} 
\def\eqstar{~{\buildrel * \over =}~} 
\def\eqques{~{\buildrel ? \over =}~} 
\def\eqsurface{~{\buildrel^{\,_{_{_\nabla}}} \over =}~} 

\def\lhs{({\rm LHS})} 
\def\rhs{({\rm RHS})} 
\def\lhsof#1{({\rm LHS~of~({#1})})} 
\def\rhsof#1{({\rm RHS~of~({#1})})} 

\def\binomial#1#2{\left(\,{\buildrel 
{\raise4pt\hbox{$\displaystyle{#1}$}}\over 
{\raise-6pt\hbox{$\displaystyle{#2}$}}}\,\right)} 

\def\Dsl{{}D \!\!\!\! /{\,}} 
\def\doubletilde#1{{}{\buildrel{\mkern1mu_\approx\mkern-1mu}%
\over{#1}}{}}

\def\hata{{\hat a}} \def\hatb{{\hat b}} 
\def\hatc{{\hat c}} \def\hatd{{\hat d}} 
\def\hate{{\hat e}} \def\hatf{{\hat f}} 

\def\circnum#1{{\ooalign%
{\hfil\raise-.12ex\hbox{#1}\hfil\crcr\mathhexbox20D}}}

\def\Christoffel#1#2#3{\Big\{ {\raise-2pt\hbox{${\scst #1}$} 
\atop{\raise4pt\hbox{${\scst#2~ #3}$} }} \Big\} }  


 
\font\smallcmr=cmr6 scaled \magstep2 
\font\smallsmallcmr=cmr5 scaled \magstep 1 
\font\largetitle=cmr17 scaled \magstep1 
\font\LargeLarge=cmr17 scaled \magstep5 
\font\largelarge=cmr12 scaled \magstep0

\def\alephnull{\aleph_0}
\def\sqrtoneovertwopi{\frac1{\sqrt{2\pi}}\,} 
\def\twopi{2\pi} 
\def\sqrttwopi{\sqrt{\twopi}} 

\def\rmA{{\rm A}} \def\rmB{{\rm B}} \def\rmC{{\rm C}} 
\def\HatC{\Hat C}

\def\alpr{\a{\hskip 1.2pt}'} 
\def\dim#1{\hbox{dim}\,{#1}} 
\def\leftarrowoverdel{{\buildrel\leftarrow\over\partial}} 
\def\rightarrowoverdel{{\buildrel\rightarrow\over%
\partial}} 
\def\ee{{\hskip 0.6pt}e{\hskip 0.6pt}} 

\def\neq{\not=} 
\def\lowlow#1{\hskip0.01in{\raise -7pt%
\hbox{${\hskip1.0pt} \!_{#1}$}}} 
\def\eqnabla{{~\, }\raise7pt\hbox{${\scriptstyle\nabla}$}{\hskip -11.5pt}={}} 

\def\atmp#1#2#3{Adv.~Theor.~Math.~Phys.~{\bf{#1}}  
(19{#2}) {#3}} 

\font\smallcmr=cmr6 scaled \magstep2 

\def\fracmm#1#2{{{#1}\over{#2}}} 
\def\fracms#1#2{{{\small{#1}}\over{\small{#2}}}} 
\def\low#1{{\raise -3pt\hbox{${\hskip 1.0pt}\!_{#1}$}}} 
\def\medlow#1{{\raise -1.5pt\hbox{${\hskip 1.0pt}\!_{#1}$}}}

\def\mplanck{M\low{\rm P}} 
\def\mplancktwo{M_{\rm P}^2} 
\def\mplanckthree{M_{\rm P}^3} 
\def\mplanckfour{M_{\rm P}^4} 
\def\mweylon{M\low S}  
\def\mhiggs{M_\medlow H}
\def\mwboson{M \low{\rm W}} 

\def\ip{{=\!\!\! \mid}} 
\def\Lslash{${\rm L}{\!\!\!\! /}\, $} 

\def\leapprox{~\raise 3pt \hbox{$<$} \hskip-9pt \raise -3pt \hbox{$\sim$}~} 
\def\geapprox{~\raise 3pt \hbox{$>$} \hskip-9pt \raise -3pt \hbox{$\sim$}~} 

\def\fR{f (R ) }
\def\FR{F \[ R \]} 
\def\FLaginv{F \[ e^{-1} \Lag_{\rm inv} \]}  
\def\LagSG{\Lag_{\rm SG}} 
\def\Laginv{\Lag_{\rm inv}} 
\def\Lagtot{\Lag_{\rm tot}} 
\def\FprimeLaginv{F\, ' \[e^{-1} \Lag_{\rm inv} \] }   
\def\FdoubleprimeLaginv{F\, '' \[e^{-1} \Lag_{\rm inv} \] }  
\def\Fzeroprime{F\, '\!\!\!_0\,} 

\def\qed{(\hbox{\it Q.E.D.})}  

\def\sqrttwo{{\sqrt 2}}  

\def\squarebrackets#1{\left[ \, {#1} \, \right]}  

\def\Biglbracket{\raise0.1pt\hbox{\Big[}{\hskip -4.6pt}\Big[\,}
\def\Bigrbracket{\,\raise0.1pt\hbox{\Big]}{\hskip -4.6pt}\Big]} 



\def\nrthreedimcont{H.~Nishino and S.~Rajpoot, \prn{75}{07}{125016}.}  

\def\kuzmincont{
S.V.~Kuzmin and D.G.C.~McKeon
\mpln{17}{02}{2605}, \hepth{0211166}.}  

\def\feldmanetalcont{See, e.g., D. Feldman, 
Z. Liu, and P. Nath, \prln{97}{06}{021801}.}  

\def\korsnathnonsusycont{B.~Kors and P.~Nath, 
\pln{586}{04}{366}, \hepph{0402047}.} 

\def\korsnathmssmcont{B.~Kors and P.~Nath, 
\jhepn{0412}{04}{005}, \hepph{0406167};
\jhepn{0507}{05}{069}, \hepph{0503208}.}    

\def\kuzminreviewcont{{\it For reviews for supersymmetric U(1) 
Stueckelberg formalism, see, e.g.,} S.V.~Kuzmin, 
{\it `Stuckelberg Formalism in Gauge and Supergauge Models'}, 
UMI-NQ-96699.}

\def\wbcont{J.~Wess and J.~Bagger, 
{\it `Superspace and Supergravity'}, 
Princeton University Press (1992).}

\def\nonabelianproblemcont{J.M.~Kunimasa and T. Goto, \ptp{37}{67}{452}; 
A.A.~Slavnov, Theor.~Math.~Phys.~{\bf 10} (1972) 99;
M.J.G.~Veltman, \np{7}{68}{637}; 
A.A.~Slavnov and L.D.~Faddeev, Theor.~Math.
\newline Phys.~{\bf 3} (1970) 312; 
A.I.~Vainshtein and I.B.~Khriplovich, Yad.~Fiz.~13 (1971) 198; 
K.I.~Shizuya, \np{121}{77}{125}; 
Y.N.~Kafiev, \np{201}{82}{341}.    
}  

\def\reviewscont{{\it For reviews, see, e.g.}, H. Ruegg and M. Ruiz-Altaba, Int.
J. Mod. Phys. A 19, 3265 (2004).}  

\def\ggrscont{S.J.~Gates, Jr., M.T.~Grisaru, M.~Ro\v cek and W.~Siegel, 
{\it `Superspace or One Thousand and One Lessons in Supersymmetry'}, 
Front.~Phys.~{\bf 58} (1983) 1-548, hep-th/0108200.} 

\def\marcusschwarzcont{N.~Marcus and J.H.~Schwarz, 
\pl{115}{82}{111}.}

\def\typeiibcont{J.H.~Schwarz, \np{226}{83}{269}.}

\def\pastietalcont{P.~Pasti, D.~Sorokin and M.~Tonin, 
\pr{52}{95}{R2447}; \pr{55}{95}{6292}.}

\def\nrnatcont{H.~Nishino and S.~Rajpoot, 
{\it `N=1 Non-Abelian Tensor Multiplet in Four Dimensions'}, 
preprint, CSULB-PA-11-03 (July, 2011).} 

\def\nrgeneralpformcont{H.~Nishino and S.~Rajpoot, 
preprint, CSULB-PA-11-05 (Oct.~2011), 
{\it `Poincare Duality with Non-Abelian Tensors in 
Arbitrary Dimensions'}.}  

\def\nrsdcont{H.~Nishino and S.~Rajpoot, 
\npn{863}{12}{510}, \hepth{1206.6175}.}  

\def\nrsusystueckelbergcont{H.~Nishino and S.~Rajpoot, 
\prn{74}{06}{105001}.}

\def\nambugoldstonecont{Y.~Nambu,
Physical Review {\bf 117} (1960) 648;
J.~Goldstone, 
Nuovo Cimento {\bf 19} (1961) 154; 
J.~Goldstone, A.~Salam and S.~Weinberg, 
Physical Review {\bf 127} (1962) 965.}  
 
\def\higgscont{P.W.~Higgs, 
Phys.~Lett.~{\bf 12} (1964) 132.}  


\def\superhiggscont{P.~Fayet, 
\np{90}{75}{104};
\nc{A31}{76}{626}.}  

\def\stueckelbergcont{A.~Proca, J.~Phys.~Radium {\bf 7} (1936) 347; 
E.C.G.~Stueckelberg, Helv.~Phys.~Acta {\bf 11}  (1938) 225;
{\it See, e.g.}, D.~Feldman, Z.~Liu and P.~Nath, \prl{97}{86}{021801}.  
{\it For reviews, see, e.g.}, H.~Ruegg and M.~Ruiz-Altaba, \ijmpn{19}{04}{3265}.} 

\def\nogocont{M.~Henneaux, V.E.~Lemes, C.A.~Sasaki, S.P.~Sorella, 
O.S.~Ventura and L.C.~Vilar, \pl{410}{97}{195}.}  

\def\ksmcont{S.~Krishna, A.~Shukla and R.P.~Malik, 
\ijmpn{26}{11}{4419}, \arXive{1008.2649}.} 

\def\bdwsncont{B.~de Wit and H.~Samtleben, 
Fortsch.~Phys.~{\bf 53} (2005) 442, hep-th/0501243; 
B.~de Wit, H.~Nicolai and H.~Samtleben, 
JHEP 0802:044,2008, \arXive{0801.1294}.}  

\def\dzcont{S.~Deser and B.~Zumino, \pl{62}{76}{335}.}  

\def\susyoriginalcont{J.L.~Gervais and B.~Sakita, 
\np{34}{71}{632}; 
Y.A.~Golfand and E.P.~Likhtman, 
JETP Lett.~{\bf 13} (1971) 323, Pisma Zh.~Eksp.~Teor.~Fiz.~{\bf 13} 
(1971) 452; 
D.V.~Volkov and V.P.~Akulov, 
JETP Lett.~{\bf 16} (1972) 438, Pisma Zh.~Eksp.~Teor.~Fiz.~{\bf 16} 
(1972) 621; 
J.~Wess and B.~Zumino, \np{70}{74}{39}.  }

\def\gswcont{M.B.~Green, J.H.~Schwarz and E.~Witten, 
{\it `Superstring Theory'}, Vols.~I \& II, 
Cambridge Univ.~Press (1986).} 

\def\mtheorycont{C.~Hull and P.K.~Townsend,
\np{438}{95}{109}; E.~Witten, \np{443}{95}{85}; 
P.K.~Townsend, {\it `Four Lectures on M-Theory'}, in {\it
`Proceedings of ICTP Summer School on High Energy
Physics and Cosmology'}, Trieste (June 1996),
hep-th/9612121;  {\it `M-theory from its Superalgebra'}, Cargese Lectures, 
1997, hep-th/9712004; T.~Banks, W.~Fischler, S.H.~Shenker
and L.~Susskind, \pr{55}{97}{5112}; 
K.~Becker, M.~Becker and J.H.~Schwarz, 
{\it `String Theory and M-Theory:  A Modern Introduction'}, 
Cambridge University Press, 2007.} 

\def\dbicont{M.~Born and L.~Infeld, Proc.~Roy.~Soc.~Lond.~%
{\bf A143} (1934) 410; {\it ibid.}~{\bf A144} (1934) 425;
P.A.M.~Dirac, Proc.~Roy.~Soc.~Lond.~{\bf A268} (1962) 57.}  

\def\gzcont{M.K.~Gaillard and B.~Zumino, 
\np{193}{81}{221}.}  

\def\bracecont{G.W.~Gibbons and D.A.~Rasheed
\np{454}{95}{185}, \hepth{9506035};  
D.~Brace, B.~Morariu and B.~Zumino, 
In *Shifman, M.A.~(ed.)~{\it `The many faces of the superworld'}, 
pp.~103-110, \hepth{9905218}; 
M.~Hatsuda, K.~Kamimura and S.~Sekiya, 
Nucl.~Phys.~{\bf 561} (1999) 341; 
P.~Aschieri, \ijmpn{14}{00}{2287}.}  

\def\kuzenkocont{S.~Kuzenko and S.~Theisen, 
\jhepn{03}{00}{034}.}   

\def\pvncont{P.~van Nieuwenhuizen, \prep{68}{81}{189}.}  

\def\schwarzsencont{J.H.~Schwarz and A.~Sen, \np{411}{94}{35}, 
\hepth{9304154}.}  

\def\shapereetalcont{A.D.~Shapere, S.~Trivedi and F.~Wilczek, 
\mpl{6}{91}{2677};
A.~Sen, 
\np{404}{93}{109}.}  
 
\def\aschierietalcont{
P.~Aschieri, D.~Brace, B.~Morariu and B.~Zumino
\npn{574}{00}{551}, \hepth{9909021}.} 
 
\def\nrtencont{H.~Nishino and S.~Rajpoot, 
\prn{71}{05}{085011}.} 

\def\sevenformcont{H.~Nicolai, P.K.~Townsend and P.~van Nieuwenhuizen, 
Lett.~Nuov.~Cim.~{\bf 30} (1981) 315; 
R.~D'Auria and P. Fr\' e, \np{201}{82}{101}.}  

\def\branecont{P.K.~Townsend, 
{\it `p-Brane Democracy'}, hep-th/9507048; 
H.~Nishino, 
\mpl{14}{99}{977}, \hepth{9802009}.}  

\def\bbscont{I.A.~Bandos, N.~Berkovits and D.P.~Sorokin, 
\np{522}{98}{214}, \hepth{9711055}.}

\def\sswcont{H.~Samtleben, E.~ Sezgin and R.~Wimmer, 
\jhepn{1112}{11}{062}.} 

\def\chucont{Chong-Sun Chu, 
DCPT-11-43, \arXive{1108.5131}.} 

\doit0{
\def\bewnscont{B.~de Wit, H.~Nicolai and H.~Samtleben, 
JHEP 0802:044,2008, \arXive{0801.1294}.}  
}

\def\nrthreecont{H.~Nishino and S.~Rajpoot, 
\prn{82}{10}{087701}.}  

\def\ntcont{H.~Nicolai and P.K.~Townsend, \pl{98}{81}{257}.}   

\def\englertwindeycont{F.~Englert and P.~Windey, 
\pr{14}{76}{2728}.} 

\def\montonenolivecont{C.~Montonen and D.I.~Olive, 
\pl{72}{77}{117}.} 

\def\olivewittencont{D.I.~Olive and E.~Witten,
\pl{78}{78}{97}.} 

\def\osborncont{H.~Osborn, 
\pl{83}{79}{321}.} 

\def\cjscont{E.~Cremmer, B.~Julia and J.~Scherk, \pl{76}{78}{409};
E.~Cremmer and B.~Julia, \pl{80}{78}{48}; \np{159}{79}{141}.}

\def\pstcont{P.~Pasti, D.P. Sorokin, M.~Tonin, 
\pl{352}{95}{59}, \hepth{9503182}.}     

\def\htwcont{C.~Hull, P.K.~Townsend, \np{438}{95}{109}; 
E.~Witten, \np{443}{95}{85}.}  

\def\tseytlinetalcont{A.A.~Tseytlin, \np{469 }{96}{51}; 
Y.~Igarashi, K.~Itoh and K.~Kamimura, \np{536 }{99}{469}.}  

\def\sstcont{M.B.~Green, J.H.~Schwarz and E.~Witten, 
{\it `Superstring Theory'}, Vols.~I \& II, 
Cambridge Univ.~Press (1986); 
K.~Becker, M.~Becker and J.H.~Schwarz, 
{\it `String Theory and M-Theory:  A Modern Introduction'}, 
Cambridge University Press, 2007.} 

\def\nogocont{M.~Henneaux, V.E.~Lemes, C.A.~Sasaki, S.P.~Sorella, 
O.S.~Ventura and L.C.~Vilar, \pl{410}{97}{195}.}  

\def\ftcont{D.Z.~Freedman, P.K.~Townsend, \np{177}{81}{282};
{\it See also}, V.I.~Ogievetsky and I.V.~Polubarinov, 
Sov.~J.~Nucl.~Phys.~{\bf 4} (1967) 156} 

\def\topologicalcont{J.~Thierry-Mieg and L.~Baulieu, \np{228}{83}{259}; 
A.H.~Diaz, \pl{203}{88}{408}; 
T.J.~Allen, M.J.~Bowick and A.~Lahiri, \mpl{6}{91}{559};
A.~Lahiri, \pr{55}{97}{5045};
E.~Harikumar, A.~Lahiri and M.~Sivakumar, \prn{63}{01}{10520}.}

\def\ferraracont{G.~Dall'Agata and S.~Ferrara, 
\npn{717}{05}{223}, \hepth{0502066}; 
G.~Dall'Agata, R.~D'Auria and S.~Ferrara, 
\pln{619}{05}{149}, \hepth{0503122}; 
R.~D'Auria and S.~Ferrara, 
\npn{732}{06}{389}, \hepth{0504108}; 
R.~D'Auria, S.~Ferrara and M.~Trigiante, 
\jhep{0509}{05}{035}, \hepth{0507225}.}  

\def\nrnacont{H.~Nishino and S.~Rajpoot, 
\hepth{0508076}, \prn{72}{05}{085020}.} 

\def\finncont{K.~Furuta, T.~Inami, H.~Nakajima and M.~Nitta, 
\ptpn{106}{01}{851}, \hepth{0106183}.} 
 
\def\scherkschwarzcont{J.~Scherk and J.H.~Schwarz, \np{153}{79}{61}.}     
 
\def\nepomechiecont{R.I.~Nepomechie, \np{212}{83}{301}.} 

\def\problemnonabeliancont{J.M.~Kunimasa and T. Goto, \ptp{37}{67}{452}; 
A.A.~Slavnov, Theor.~Math.~Phys.~{\bf 10} (1972) 99;
M.J.G.~Veltman, \np{7}{68}{637}; 
A.A.~Slavnov and L.D.~Faddeev, Theor.~Math.
\newline Phys.~{\bf 3} (1970) 312; 
A.I.~Vainshtein and I.B.~Khriplovich, Yad.~Fiz.~13 (1971) 198; 
K.I.~Shizuya, \np{121}{77}{125}; 
Y.N.~Kafiev, \np{201}{82}{341}.    
}

\def\aflcont{L.~Andrianopoli, S.~Ferrara and M.A.~Lledo, 
\hepth{0402142}, \jhep{0404}{04}{005};  
R.~D'Auria, S.~Ferrara, M.~Trigiante and S.~Vaula, 
\pln{610}{05}{270}, \hepth{0412063}.}   

\def\mt{M.~Blau and G.~Thompson,
\ap{205}{91}{130}.}  

\def\hk{M.~Henneaux and B.~Knaepen, 
\pr{56}{97}{6076}, \hepth{9706119}.}

\def\originalcont{S.~Ferrara, B.~Zumino, and J.~Wess, \pl{51}{74}{239}; 
W.~Siegel, \pl{85}{79}{333}; U.~Lindstrom and M.~Ro\v cek, 
\np{222}{83}{285}; 
{\it For reviews of linear multiplet coupled to SG, see, e.g}., P.~Bine«truy, 
G.~Girardi, and R.~Grimm, \prn{343}{01}{255}, {\it and
references therein}.} 

\def\stringrelatedcont{S.~Ferrara and M.~Villasante, \pl{186}{87}{85};
P.~Bin\' etruy, G.~Girardi, R.~Grimm, and M.~Muller, \pl{195}{87}{389}; 
S.~Cecotti, S.~Ferrara, and M.~Villasante, Int.~Jour.~Mod.~Phys.
\newline {\bf A2} (1987) 1839; 
M.K.~Gaillard and T.R.~Taylor, \np{381}{92}{577};
V.S.~Kaplunovsky and J.~Louis, \np{444}{95}{191};
P.~Bin\' etruy, F.~Pillon, G.~Girardi and R.~Grimm, \np{477}{96}{175}; 
P.~Bin\' etruy, M.K.~Gaillard and Y.-Y.~Wu, \pl{412}{97}{288}; 
\np{493}{97}{27}493, \ibid{B481}{96}{109};
D.~Lu¬st, S.~Theisen and G.~Zoupanos, \np{296}{88}{800}; 
J.~Lauer, D.~L\" ust and S.~Theisen, \np{304}{88}{236}.}  

\def\threealgebracont{N.~Lambert and C.~Papageorgakis, 
\jhepn{08}{10}{083}; 
K-W.~Huang and W-H.~Huang, 
arXiv:{1008.3834} \newline [hep-th]; 
S.~Kawamoto, T.~Takimi and D.~Tomino, 
J.~Phys.~{\bf A44} (2011) 325402, \arXive{1103.1223};
Y.~Honma, M.~Ogawa, S.~Shiba, 
\jhepn{1104}{11}{2011}, \arXive{1103.1327}; 
C.~Papageorgakis and C.~Saemann, \jhep{1105}{11}{099}, 
\arXive{1103.6192}.}  

\def\bwcont{J.~Wess and J.~Bagger, {\it `Superspace and Supergravity'}, 
Princeton University Press (1992).}  
 
\def\nrhigherdimcont{H.~Nishino and S.~Rajpoot, 
{\it `Poincar\' e Duality with Non-Abelian Tensors in Arbitrary Dimensions'}, 
CSULB-PA-11-4.}  

\def\cllcont{T.E.~Clark, C.H.~Lee and S.T.~Love, 
\mpl{4}{89}{1343}.}  

\def\khelashvilicont{G.A.~Khelashvili and V.I.~Ogievetsky, 
\mpl{6}{91}{2143}.}

\def\buchbindercont{I.L.~Buchbinder and N.G.~Pletnev, 
Theor.~Math.~Phys.~{\bf 157}  (2008) 1383, \arXive{0810.1583}.} 

\def\swcont{E.~Sezgin and L.~Wulff, \jhepn{1303}{13}{023}, \arXive{1212.3025s}.}    

\def\nrdilatonaxioncont{H.~Nishino and S.~Rajpoot, 
\prn{76}{07}{065004}.} 

\def\nrnatcont{ H. Nishino and S. Rajpoot, 
\arXive{1204.1379}, \prn{85}{12}{105017}.}

\def\gatespformcont{S.J.~Gates Jr., 
\np{184}{81}{381}.}  

\def\buchbindercont{I.L.~Buchbinder and S.M.~Kuzenko, 
{\it `Ideas and Methods of Supersymmetry and
Supergravity or a Walk Through Superspace'}, Bristol, UK: IOP (1998).}  

\def\mullercont{M.~Muller, 
{\it `Consistent Classical Supergravity Theories'}, 
Lecture Notes in Physics {\bf 336}, Springer, Berlin (1989) 1; 
CERN-TH-4984-88; 
P.~Binetruy, G.~Girardi and R.~Grimm, 
\prepn{343}{01}{255}.}  

\def\superfieldcont{P.~Binetruy, G.~Girardi and R.~Grimm, 
\prepn{343}{01}{255}.}  

\def\nrstueckelcont{H.~Nishino and S.~Rajpoot, 
\npn{872}{13}{213}, \arXive{0806.0660}.}  

\def\higgsbosoncont{ATLAS Collaboration, \pln{716}{12}{1}, 
\arXive{1207.7214}; CMS Collaboration, \pln{716}{12}{30}, 
\arXive{1207.7235}.} 

\def\thetavacuumcont{
A.A.~Belavin, A.M.~Polyakov, A.S.~Schwartz, and Yu.S.~Tyupkin, 
\pl{59}{75}{85}; 
G.~Õt Hooft, \prl{37}{76}{8}; \pr{14}{76}{3432}, \ibid{18}{78}{2199(E)}.}  

\def\uonecont{E.~Witten, \np{156}{79}{269}; 
G.~Veneziano, \np{159}{79}{213}.}



\doit0{
\def\framing#1{\doit{#1}  {\framingfonts{#1} 
\border\headpic  }} 
%
\framing{0} 
} 

\def\Cases#1{\left \{ \matrix{\displaystyle #1} \right.}   

\def\fIJK{f^{I J K}} 

\doit0{
\def\matrix#1{\null\ , \vcenter{\normalbaselines\m@th
	\ialign{\hfil$##$\hfil&&\quad\hfil$##$\hfil\crcr 
	  \mathstrut\crcr\noalign{\kern-\baselineskip}
	  #1\crcr\mathstrut\crcr\noalign{\kern-\baselineskip}}}\ ,} 
} 

\def\ialign{\everycr={}\tabskip=0pt \halign} 

\doit0{
\def\matrixs#1{\null\ , {\normalbaselines \m@th
	\ialign{\hfil$##$\hfil&&\quad\hfil$##$\hfil\crcr 
	  \mathstrut\crcr\noalign{\kern-\baselineskip}
	  #1\crcr\mathstrut\crcr\noalign{\kern-\baselineskip}}}\ ,} 
} 



\doit0{
\vskip -0.6in 
{\bf Preliminary Version (FOR YOUR EYES
ONLY!)\hfill\today} \\[-0.25in] 
\\[-0.0in]  
}
\vskip -0.3in  
\doit0{
{\hbox to\hsize{\hfill
hep-th/yymmnnn}} 
} 
\doit0{\vskip 0.1in  
{\hbox to\hsize{\hfill CSULB--PA--13--2}}  
\vskip -0.05in 
{\hbox to\hsize{\hfill 
}} 
}  

~~~ 
\vskip 0.15in 

\begin{center}  

{\Large\bf N {\hskip 3pt}= {\hskip 3pt}1~ 
Supersymmetric} \\ 
{\Large\bf Non-Abelian Compensator Mechanism} \\ 
\smallskip 
{\Large\bf for Extra Vector Multiplet} \\
\vskip 0.05in 

\baselineskip 9pt 

\vskip 0.26in 

Hitoshi ~N{\smallcmr ISHINO}%
\footnotes{E-Mail: H.Nishino@csulb.edu} ~and
~Subhash ~R{\smallcmr AJPOOT}%
\footnotes{E-Mail: Subhash.Rajpoot@csulb.edu} 
\\[.16in]  {\it Department of Physics \& Astronomy}
\\ [.015in] 
{\it California State University} \\ [.015in]  
{\it 1250 Bellflower Boulevard} \\ [.015in]  
{\it Long Beach, CA 90840} \\ [0.02in] 

\vskip 1.8in 

{\bf Abstract}\\[.1in]  
\end{center}  
\vskip 0.1in 

\baselineskip 14pt 

~~~We present a variant formulation of 
$~N=1$~ supersymmetric compensator mechanism for an arbitrary 
non-Abelian group in four dimensions.  This formulation resembles our 
previous variant supersymmetric compensator mechanism in 4D.  
Our field content consists 
of the three multiplets:  (i) A Non-Abelian Yang-Mills multiplet 
$~(A\du\m I, \l^ I, C\du{\m\n\r} I )$, (ii) a tensor multiplet $~(B\du{\m\n} I, 
\chi^I, \varphi^I)$~ and an extra vector multiplet $~(K\du\m I, \r^I, C\du{\m\n\r} I)$~   
with the index $~{\scst I}$~~for the adjoint representation of a non-Abelian 
gauge group.  The $~C\du{\m\n\r} I$~ is originally an auxiliary field dual to the conventional auxiliary field $~D^I$~ for the extra vector multiplet. 
The vector $~K\du\m I$~ and the tensor $~C\du{\m\n\r} I$~
get massive, after absorbing respectively the scalar $~\varphi^I$~ and the tensor $~B\du{\m\n}I$.  The superpartner fermion $~\r^I$~ acquires a Dirac mass shared with $~\chi^I$.  We fix all non-trivial cubic interactions in the total lagrangian, 
all quadratic terms in supersymmetry transformations, 
and all quadratic interactions in field equations.  The action 
invariance and the super-covariance of all field equations are confirmed up 
to the corresponding orders.

\vskip 0.5in  

\baselineskip 8pt 
\leftline{\small PACS:  11.15.-q, 11.30.Pb, 12.60.Jv}  
\vskip 0.03in 
\leftline{\small Key Words: \hfill N=1$\,$ Supersymmetry, \hfill 
Proca-Stueckelberg Formalism, \hfill Non-Abelian Group,} 
\leftline{\small  {\hskip 0.87in} Compensators, \hfill Vector Multiplet, 
\hfill Tensor Multiplet, \hfill Non-Abelian Tensors.} 


\vfill\eject  

\baselineskip 18pt 

\oddsidemargin=0.03in 
\evensidemargin=0.01in 
\hsize=6.5in
\topskip 0.16in 
\textwidth=6.5in 
\textheight=9in 
\flushbottom
\footnotesep=1.0em
\footskip=0.36in 
\def\baselinestretch{1.0} 

\def\fixedpoint{20.0pt} 
\baselineskip\fixedpoint    

\pageno=2 



\centerline{\bf 1.~~Introduction}  

Recently, there have been considerable developments  
\ref\sw{\swcont}%
\ref\nrstueckel{\nrstueckelcont}   
for the supersymmetrization of the Proca-Stueckelberg compensator mechanism 
\ref\stueckelberg{\stueckelbergcont}.   
The supersymmetrization of {\it non-Abelian} 
compensator mechanism was first performed in late 1980's 
\ref\cll{\cllcont}.
{\it Abelian} supersymmetric Proca-Stueckelberg mechanism  
\ref\buchbinder{\buchbindercont}  
has a direct application to MSSM 
\ref\korsnathmssm{\korsnathmssmcont}.  
In \sw, general representations of {\it non-Abelian} group are analyzed, 
and higher-order terms have been also fixed.  
Even though the original Higgs mechanism 
\ref\higgs{\higgscont} 
has been established experimentally
\ref\higgsboson{\higgsbosoncont}, 
the Proca-Stueckelberg type compensator mechanism 
for massive gauge fields \stueckelberg\ is still an important theoretical alternative.    

In our recent paper \nrstueckel, we presented a {\it variant} supersymmetric 
compensator mechanism, both in component and superspace 
\ref\wb{\wbcont}, 
with a field content different from \cll.  
Our formulation in \nrstueckel\ differs also from \sw.   
The field content in \nrstueckel\ consists of two multiplets:  
Yang-Mills (YM) vector multiplet (VM) $~(A\du\m I, \l^I, C\du{\m\n\r} I)$, 
and the tensor multiplet (TM) $~(B\du{\m\n} I , \chi^I , \varphi^I)$.  
The $~C\du{\m\n\r} I \-$field is dual to the conventional auxiliary field $~D^I$.  
The `dilaton' $~\varphi^I$~ (or $~B\du{\m\n} I $) 
is absorbed into the longitudinal component of $~A\du\m I$~ (or $~C\du{\m\n\r} I$), making the latter massive \nrstueckel.  This compensation mechanism works even with $~C\du{\m\n\r}I$~ in the adjoint representation.  

In this present paper, we demonstrate yet another field content 
as a supersymmetric compensator system in which 
an extra vector in the adjoint representation absorbs a scalar.  
We have three multiplets 
VM $~(A\du\m I, \l^I)$, TM $~(B\du{\m\n} I , \chi^I, \varphi^I)$~, and 
the extra vector multiplet (EVM) $~(K\du\m I, \r^I , C\du{\m\n\r} I)$.  
The $~\varphi^I$~ and $~B\du{\m\n} I$~ in the TM 
are compensator fields, respectively absorbed into 
$~K\du\m I$~ and $~C\du{\m\n\r} I\-$fields in the EVM.  
{\it Before} the absorptions, the {\it on-shell} degrees of freedom (DOF) count as 
$~A\du\m I (2), ~\l^I (2);~ B\du{\m\n} I (1),~ \chi^I (2) , ~\varphi^I(1); 
~ K\du\m I (2), ~\r^I (2), ~C\du{\m\n\r} I (0)$.    
{\it After} the absorption, the {\it on-shell} DOF count as 
$~A\du\m I (2), ~\l^I (2); ~K\du\m I (3), ~\r^I (4), ~ 
C\du{\m\n\r} I (1)$, as summarized in the following Table.       

\vskip 0.05in

\vbox{ 
\oddsidemargin=3.0in 
\evensidemargin=0.0in 
\hsize=6.5in 
\textwidth=5.5in 
\textheight=9in 
\flushbottom 
\footnotesep=1.0em 
\footskip=0.36in 
\def\baselinestretch{0.8} 
%
\begin{center}
\begin{tabular}{|c|c|c|c|c|c|c|c|c|c|c|c|c|} 
\noalign {\vskip -0.00cm} 
\hline 
\noalign {\vskip 0.03cm} 
{\largelarge DOF before Absorptions} &{\hskip -10pt} & $A\du\m I$ & $\l^I$ & 
{\hskip -10pt} & $B\du{\m\n}I $ & $ \chi^I$ & $\varphi^I $ 
& {\hskip -10pt} & $K\du\m I $ & $\r^I $ & $C\du{\m\n\r} I$ \\ 
\hline 
\noalign{\vskip 0.03cm}  
\hline 
\noalign {\vskip 0.03cm} 
{\largelarge On-Shell} & {\hskip -10pt} & 2 & 2 & {\hskip -10pt} & 1 & 2 & 1  
& {\hskip -10pt} & 2 & 2 & 0\\ 
\hline 
\noalign {\vskip 0.03cm} 
{\largelarge Off-Shell} &{\hskip -10pt} &  3 & 4 & {\hskip -10pt} & 3 & 4 & 1  
& {\hskip -10pt} & 3 & 4 & 1 \\ 
\hline
\noalign {\vskip 0.03cm} 
\noalign {\vskip 0.5cm} 
\hline 
{\largelarge DOF after Absorptions} & {\hskip -10pt}  & 
$A\du\m I$ & $\l^I$  & 
{\hskip -10pt} & $B\du{\m\n}I $ & $ \chi^I$ & $\varphi^I $  
& {\hskip -10pt} & $K\du\m I $ & $\r^I $ & $C\du{\m\n\r} I$ \\ 
\hline 
\noalign{\vskip 0.03cm}  
\hline
\noalign {\vskip 0.03cm} 
{\largelarge On-Shell} &{\hskip -10pt} & 2 & 2 & {\hskip -10pt} & 0 & 0 & 0  
& {\hskip -10pt} & 3 & 4 & 1 \\ 
\hline 
\noalign {\vskip 0.03cm} 
{\largelarge Off-Shell} & {\hskip -10pt} & 3 & 4  &{\hskip -10pt}  & 0 & 0 & 0 
& {\hskip -10pt} & 6 & 8 & 2 \\ 
\hline
\end{tabular} 
\vskip 0.16in
{\largelarge Table 1:  ~DOF of Our Field Content} 
\end{center} 
\vspace{-0.3cm} 
} 

\oddsidemargin=0.03in 
\evensidemargin=0.01in 
\hsize=6.5in
\topskip 0.32in 
\textwidth=6.5in 
\textheight=9in 
\flushbottom
\footnotesep=1.0em
\footskip=0.36in 
\def\baselinestretch{0.8} 


Our new system differs from our recent work 
\nrstueckel\ in terms of the three aspects:
\vskip -0.07in 

\Item{\bf (i)} Our present system has three multiplets VM, TM and EVM, 
while that in \nrstueckel\ has only VM and TM.  
The new multiplet is EVM $~(K\du\m I, \r^I , C\du{\m\n\r} I)$, 
where $~K\du\m I$~ (or $~C\du{\m\n\r} I$) absorbs 
~$\varphi^I$~ (or $~B\du{\m\n} I$), getting massive.     

\vskip -0.05in 

\Item{\bf (ii)} 
The vector field getting massive is {\it not} $~A\du\m I$, 
but is the extra vector field $~K\du\m I$.  

\vskip -0.05in 

\Item{\bf (iii)} The VM $~(A\du\m I, \l^I)$~ has {\it no} auxiliary field, 
while the EVM has the auxiliary field $~C\du{\m\n\r} I$.
So our VM is {\it on-shell}, while our TM and EVM are {\it off-shell}.

\bigskip\bigskip 



\centerline{\bf 2.~~Field Strengths and Tensorial Transformations} 
\nobreak 

The field strengths for our bosonic fields $~A\du\m I, ~B\du{\m\n\r} I, 
~C\du{\m\n\r} I, ~K\du\m I$~ and $~\varphi^I$~ are respectively 
$$ \li{ F\du{\m\n} I \equiv & + 2 \partial_{\[ \m} A\du{\n \]} I 
				+ m \fIJK A\du\m J A\du\n K ~~, 
&(2.1\rma) \cr 
\calG\du{\m\n\r} I \equiv & + 3 D_{\[ \m} B\du{\n\r\] } I 
		+ m \, C\du{\m\n\r} I 
		- 3 m^{-1} \fIJK F\du{\[\m\n}J \calD_{ \r\]} \varphi^K ~~, 
&(2.1\rmb) \cr 
H\du{\m\n\r\s} I \equiv & + 4 D_{\[ \m} C\du{\n\r\s \] } I 
				+ 6 \fIJK F\du{\[ \m\n} J B\du{\r\s \]} K ~~,  
&(2.1\rmb) \cr 
L\du{\m\n} I \equiv & + 2 D_{\[ \m} K\du{\n \]} I 
		+ \fIJK F\du{\m\n} J \varphi^K ~~, 
&(2.1\rmd) \cr 
\calD_\m \varphi^I \equiv & + D_\m \varphi^I + m K\du\m I ~~, 
&(2.1\rme) \cr } $$ 
We use $~m$~ for the YM coupling constant, while $~D_\m$~ is 
the YM-covariant derivative.    
The $~\calG$~ in (2.1b) instead of $~G$~ 
is a reminder that this field strength 
has an extra term $~m^{-1} F\wedge B$.  Similarly,  
$~\calD_\m$~ in (2.1e) is used to be distinguished from $~D_\m$.   
The $~m \, C$~ and $~m \, K\-$terms in the respective field strength 
$~\calG$~ and $~\calD\varphi$~ are suggestive  
that these field strengths can be absorbed into the field redefinitions of 
$~C$~ and $~K$.  

The Bianchi identities for our field strengths are
$$ \li{ D_{\[\m} F\du{\n\r\]} I \equiv & 0 ~~, 
&(2.2\rma) \cr 
D_{\[\m} \calG \du{\n\r\s \]} I \equiv & + \fracm 14 m H\du{\m\n\r\s} I 
			- \fracm 32 \fIJK F\du{\[\m\n} J L\du{\r\s\]} K ~~, 
&(2.2\rmb) \cr 
D_{\[\m} L\du{\n\r \]} I \equiv & + \fIJK F\du{\[\m\n} J \calD_{\r \]} 
		\varphi^K ~~, 
&(2.2\rmc) \cr 
D_{\[\m} \calD_{\n\]} \varphi^I \equiv & + \fracm 12 m L\du{\m\n} I ~~, 
&(2.2\rmd) \cr } $$   

There should be proper tensorial transformations \sw\nrstueckel\ 
associated with $~B\du{\m\n} I, ~C\du{\m\n\r} I$~ and $~K\du\m I$~ 
which are symbolized as $~\d_\b, ~\d_\g$~ and $~\d_\k$.  The last $~\d_\k$~ is 
for the extra vector $~K\du\m I$~ which is also a kind of 
`tensor' in adjoint representation:   
$$ \li{ & \d_\a (A\du\m I, ~B\du{\m\n} I, ~C\du{\m\n\r} I , ~K\du\m I,~ \varphi^I) \cr 
	& ~~ = (D_\m \a^I ,  ~- \fIJK \a^J B\du{\m\n} K ,  ~ - \fIJK \a^J C\du{\m\n\r} K , 
	     ~- \fIJK \a^J K\du\m K ,  ~- \fIJK \a^J \varphi^K ) ~~,~~~~~ ~~~ 
&(2.3\rma) \cr 
& \d_\b (A\du\m I,  ~B\du{\m\n} I,  ~C\du{\m\n\r} I ,  ~K\du\m I,  ~\varphi^I) 
	= (0,  ~ + 2 D_{\[\m} \b\du{\n\]} I ,  ~- 3\fIJK F\du{\[\m\n} J \b\du{\r\]} K , 
		 ~0,  ~0 ) ~~, 
&(2.3\rmb) \cr 
& \d_\g (A\du\m I,  ~B\du{\m\n} I,  ~C\du{\m\n\r} I ,  ~K\du\m I, \varphi^I) 
	= (0,  ~- m \g_{\m\n} ,  ~+ 3 D_{\[\m} \g\du{\n\r\]} I ,  ~0,  ~0) ~~, 
&(2.3\rmc) \cr 
& \d_\k (A\du\m I,  ~B\du{\m\n} I,  ~C\du{\m\n\r} I ,  ~K\du\m I,  ~\varphi^I) 
	= (0, ~0, ~0,  ~D_\m \k^I,  ~- m \k^I) ~~, 
&(2.3\rmd) \cr }  $$ 
where $~\d_\a$~ is the standard YM gauge transformation. 

Our field strengths are all covariant under $~\d_\a$, while 
invariant under $~\d_\b, ~\d_\g, ~\d_\g$~ and $~\d_\k$: 
$$ \li{ & \d_\a (F\du{\m\n} I,  ~\calG\du{\m\n\r} I,  ~H\du{\m\n\r} I , 
	 ~L\du{\m\n} I,  ~D_\m\varphi^I) \cr 
& ~~~~~ = - \fIJK \a^J  (F\du{\m\n} K, ~\calG\du{\m\n\r} K , ~H\du{\m\n\r\s} K , ~ 
		L\du{\m\n} K,  ~D_\m\varphi^K) , ~~~~~ ~~~ 
&(2.4\rma) \cr 
& \d_\b  (F\du{\m\n} I,  ~\calG\du{\m\n\r} I,  ~H\du{\m\n\r} I , 
	 ~L\du{\m\n} I,  ~D_\m\varphi^I) 
	= (0,0,0,0,0)~~, 
&(2.4\rmb) \cr 
& \d_\g   (F\du{\m\n} I,  ~\calG\du{\m\n\r} I,  ~H\du{\m\n\r} I , 
	 ~L\du{\m\n} I,  ~D_\m\varphi^I) = (0,0,0,0,0)~~, 
&(2.4\rmc) \cr 
& \d_\k   (F\du{\m\n} I,  ~\calG\du{\m\n\r} I,  ~H\du{\m\n\r} I , 
	 ~L\du{\m\n} I,  ~D_\m\varphi^I) = (0,0,0,0,0)~~.   
&(2.4\rmd) \cr} $$ 

The transformations (2.3c) and (2.3d) indicate that 
the $~C\du{\m\n\r} I$~ and $~K\du\m I\-$fields respectively   
can absorb the compensators $~B\du{\m\n} I$~ and $~\varphi^I$.  

\bigskip\bigskip




\centerline{\bf 3.~~Lagrangian and $\,$N=1$\,$ Supersymmetry} 
\nobreak 
	 
Once the invariant field strengths $~F,~\calG,~H, ~L$~ and $~\calD\varphi$~ 
have been established, 
it is straightforward to construct a lagrangian, invariant also under 
$~N=1$~ supersymmetry.  Our Action $~I \equiv \int d^4 x \, \Lag $~ 
has the lagrangian\footnotes{We also use the symbol $~{\scst \[ r \]}$~ for 
totally antisymmetric indices $~{\scst \r_1 \cdots \r_r}$~ to save space.  
Our notation is $ (\eta_{\m\n} ) \equiv \hbox{diag.}~(-,+,+,+), ~\e^{0123} = + I, ~\e_{\m_1\cdots \m_{4-r} \[ r \]} 
\,\e^{\[ r \] \s_1\cdots\s_{4-r}} = -(-1)^r (4-r)! \, (r!)\, \d\du{\[ \m_1}{\s_1} 
\cdots\d\du{\m_{4-r} \]}{\s_{4-r}}, ~\g_5 \equiv + i \g_0 \g_1 \g_2\g_3,~ 
\e^{ \[ 4-r \]  \[ r \]} \g_{ \[ r \] } = - i (-1)^{r(r-1)/2} (r!) 
\, \g_5 \g^{\[ 4-r \] } $.}  
$$ \li{ \Lag = & - \fracm 14 (F\du{\m\n} I)^2 
			+ \fracm 12 (\Bar\l{}^I \Dsl \l^I )  
		- \fracm1{12} (\calG\du{\m\n\r} I)^2 + \fracm 12 (\Bar\chi{}^I \Dsl \chi^I ) 
		- \fracm 12 (\calD_\m\varphi)^2 
		- \fracm1{48} (H\du{\[4\]} I)^2 \cr 
& + \fracm 12 (\Bar\r{}^I \Dsl \r^I )  
	- \fracm14 (L\du{\m\n} I)^2 + m \, (\Bar\chi{}^I \r^I) 
	+ \fracm 1{48} \fIJK (\Bar\l{}^I \g^{\[ 4\]} \chi^J) H\du{\[ 4 \] } K 
		- \fracm 1{12} \fIJK (\Bar\l{}^I \g^{ \[ 3 \]} \r^J) \, 
			\calG\du{\[ 3 \] } K  \cr 
& - \fracm 14 \fIJK (\Bar\chi{}^I \g^{\m\n} \r^J) F\du{\m\n} K 
			- \fracm12 \fIJK (\Bar\l{}^I \g^\m \r^J) \calD_\m\varphi^K 
			 + \fracm 14 \fIJK (\Bar\l{}^I \g^{\m\n} \chi^J) L\du{\m\n} K 
			 ~~ ~~~~~ ~~~
&(3.1) \cr } $$  
up to quartic terms $~\order\phi4$.  
The kinetic terms of $~B$~ and $~\varphi$, 
namely, the $~(\calG\du{\m\n\r}I )^2$~ and $~(\calD_\m\varphi)^2\-$terms, 
which respectively contain $~m^2  C^2$~ and $~m^2 K^2\-$terms, 
play the role of mass terms for the $~C$~ and $~K\-$fields, after 
the absorptions of $~D B$~ by $~C$ and $~\calD\,\varphi$~ by $~K$.   
Because of $~N=1$~ supersymmetry, this compensator mechanism between 
TM and EVM works also for fermionic partners.  Namely, 
the original $~\chi^I\-$field in TM is mixed with the $~\r^I\-$field in EVM,  
forming the Dirac mass term $~m \, (\Bar\chi{}^I \r^I)$.  

The $~N=1$~ supersymmetry transformation rule of our multiplets is 
$$ \li{ \d_Q A\du\m I = & + (\Bar\e \g_\m\l^I) ~~, 
&(3.2\rma) \cr 
\d_Q \l^I = & + \fracm 12 (\g^{\m\n} \e) F\du{\m\n} I 			
			+ \fracm 12 \fIJK (\g_5 \e) (\Bar\chi{}^J \g_5 \r^K) ~~, 
&(3.2\rmb) \cr 
\d_Q B\du{\m\n} I = & + (\Bar\e\g_{\m\n} \chi^I) 
			+ 2 m^{-1} \fIJK (\Bar\e \g_{\[\m | } \l^J) \calD_{| \n\]} \varphi^K 
				- m^{-1} \fIJK (\Bar\e \chi^J) F\du{\m\n} K ~~, 		
&(3.2\rmc) \cr 
\d_Q \chi^I = & + \fracm 16 (\g^{\m\n\r} \e) \calG\du{\m\n\r} I 
			- (\g^\m \e) \calD_\m \varphi^I 
			+ \fracm 12 \fIJK (\g^\m \r^J) (\Bar\e\g_\m \l^K) \cr 
& - \fracm 12 \fIJK \r^J (\Bar\e\l^K) 
		+ \fracm 12 \fIJK (\g_5 \r^J)  (\Bar\e\g_5 \l^K) ~~, 
&(3.2\rmd) \cr 
\d_Q \varphi^I = & + (\Bar\e \chi^I) ~~, 
&(3.2\rme) \cr 
\d_Q K\du\m I = & + (\Bar\e\g_\m \r^I) 
			- \fIJK (\Bar\e \g_\m \l^J) \varphi^K ~~, 
&(3.2\rmf) \cr 
\d_Q \r^I = & + \fracm 12 (\g^{\m\n} \e) L\du{\m\n} I 
			- \fracm1{24} ( \g^{\m\n\r\s} \e) H\du{\m\n\r\s} I 
					+ \fracm 12 \fIJK (\g^\m \chi^J) (\Bar\e \g_\m\l^K) \cr 
& + \fracm 12 \fIJK \chi^J (\Bar\e\l^K ) 
	+ \fracm 12 \fIJK (\g_5 \chi^J) (\Bar\e\g_5 \l^K) ~~ , 
&(3.2\rmg) \cr 
\d_Q C\du{\m\n\r} I = & + (\Bar\e \g_{\m\n\r} \r^I) 
				- 3 \fIJK (\Bar\e \g_{\[\m | } \l^J) B\du{| \n\r\]} K ~~. 
&(3.2\rmh) \cr } $$ 

An important corollary is for the arbitrary variations of 
our field strengths:
$$ \li{ \d F\du{\m\n} I = & + 2 D_{\[\m | } (\d A\du{| \n\]} I) ~~, 
&(3.3\rma) \cr 
\d \calG\du{\m\n\r} I = & + 3 D_{\[\m | } (\Tilde\d B\du{| \n\r\]} I) 
				+ m (\Tilde \d C\du{\m\n\r} I)  \cr 
& - 3 \fIJK (\d A\du{\[\m | } J ) L\du{| \n\r \] } K 
		- 3 \fIJK F\du{\[\m\n | } J (\Tilde \d K\du{|\r\]} K ) ~~,  	
&(3.3\rmb) \cr 
\d H\du{\m\n\r\s} I = & + 4 D_{ \[\m |} (\Tilde\d C\du{| \n\r\s\]} I) 
			+ 4 \fIJK (\d A\du{ \[ \m | } J) \left( \calG\du{| \n\r\s \] } K   
			+ 3 m^{-1} f^{K L M} L\du{| \n\r|} L \calD_{| \s\]} \varphi^M \right) \cr 
& - 6 \fIJK (\Tilde \d B\du{\[\m\n |} J ) F\du{| \r\s\]} K ~~, 
&(3.3\rmc) \cr 
\d L\du{\m\n} I = & + 2 D_{\[\m | } (\Tilde\d K\du{ | \n\]} I ) 
			+ 2 \fIJK (\d A\du{\[\m |} J ) \calD_{ | \n\]} \varphi^K 
					+ \fIJK F\du{\m\n} J (\d\varphi^K) ~~,   
&(3.3\rmd) \cr  
\d (D_\m \varphi^I) = & + D_\m (\d\varphi^I ) + m (\Tilde\d K\du\m I) ~~,  
&(3.3\rme) \cr } $$ 
These results are valid up to $~\order\phi 3\-$terms.  
The modified transformations $~\Tilde\d B, ~\Tilde\d C$~ and $~\Tilde\d K$~ are 
defined by 
$$\li{ \Tilde\d B\du{\m\n} I \equiv & + \d B\du{\m\n} I
			- 2 m^{-1} \fIJK (\d A\du{\[\m | } J) \calD_{ | \n\]} \varphi^K 
			- m^{-1}\fIJK F\du{\m\n} K (\d\varphi^K) ~~, 
&(3.4\rma) \cr  
\Tilde\d C\du{\m\n\r} I \equiv & + \d C\du{\m\n\r} I 
		+ 3 \fIJK (\d A\du{\[\m|} J) B\du{|\n\r\]} K ~~, ~~~~ 
	\Tilde\d K\du\m I \equiv + \d K\du\m I  + \fIJK (\d A\du\m J) \varphi^K ~~. 
	~~~~~ ~~~~~ ~~
&(3.4\rmb) \cr  } $$ 
A special case of (3.3) is the supersymmetry transformation rule, 
$$ \li{ \d_Q F\du{\m\n} I = & - 2 (\Bar\e \g_{\[\m } D_{\n\]} \l^I) ~~, 
&(3.5\rma) \cr 
\d_Q \calG\du{\m\n\r} I = & + 3 (\Bar\e \g_{\[ \m\n} D_{\r \]} \chi^I) 
				+ m (\Bar\e\g_{\m\n\r} \r^I) \cr 
& - 3 \fIJK (\Bar\e \g_{\[\m | } \l^J) L\du{| \n\r \]} K 
			+ 3 \fIJK (\Bar\e \g_{ \[ \m |} \r^J) F\du{| \n\r\] } K ~~,  
&(3.5\rmb) \cr 
\d_Q H\du{\m\n\r\s} I = & - 4 (\Bar\e \g_{\[ \m\n\r} D_{\s\]} \r^I) 
					+ 4 \fIJK (\Bar\e \g_{\[\m |} \l^J ) \calG\du{| \n\r\s\]} K 
					- 6 \fIJK (\Bar\e \g_{\[\m\n|} \chi^J) F\du{|\r\s\]} K~~, 
					~~~~~ ~~~   
&(3.5\rmc) \cr 
\d_Q L\du{\m\n} I = & - 2 (\Bar\e \g_{\[\m} D_{\n\]} \r^I) 
		+ 2 \fIJK (\Bar\e\g_{\[\m|} \l^J) \calD_{|\n\]} \varphi^K 
		- \fIJK (\Bar\e\chi^J) F\du{\m\n} K ~~,  
&(3.5\rmd) \cr 
\d_Q (\calD_\m\varphi^I) = & + ( \Bar\e D_\m \chi^I) 
			+ m (\Bar\e \g_\m\r^I) ~~,  
&(3.5\rme) \cr } $$  
up to $~\order\phi 3\-$terms.  
In particular, there should be {\it no} `bare' potential-field terms, such as `bare' 
$~B\du{\m\n} I$~ or `bare' $~\varphi^I\-$term without derivatives in (3.3).  
The modified transformations (3.4) explain why the terms 
in $~\d_Q B\du{\m\n} I$~ (3.2c), $~\d_Q C\du{\m\n\r} I$~ (3.2h) and 
$~\d_Q K\du\m I$~ (3.2f) other than their first linear terms are required. 
In other words, all the {\it tilted} transformations $~\Tilde\d_Q B\du{\m\n} I, 
~\Tilde\d_Q C\du{\m\n\r} I, ~\Tilde\d_Q K\du\m I$ contain only the linear terms 
in (3.2c), (3.2h) and (3.2f), respectively.  

Note the peculiar $~m^{-1} F\wedge B\-$term in $~\calG$~ in (2.1b).  
The general variation of this term is 
$$ \li{ & \d \left( - 3 m^{-1} \fIJK F\du{\[\m\n} J \calD_{\r\]} \varphi^K \right) \cr  
& ~~~ = + 3 D_{\[\m | } 
	\left[ \, - 2 m^{-1} \fIJK (\d A\du {| \n | } J ) \, \calD_{| \r \]} \varphi^K \, \right] 
	- 6 m^{-1} \fIJK (\d A\du{\[ \m |} J) D_{| \n | } \calD_{| \r \]} \varphi^K ~~. 
&(3.6)  \cr } $$ 
The last term is proportional to $~(\d A) \wedge \, L $~ 
with the original $~m^{-1}$~  
cancelled by $~m$~ in the former resulting in only a $~m^0\-$term, 
interpreted as the third term in (3.3b).  The first term of (3.6) with $~m^{-1}$~ 
is absorbed into the second term of 
$~\Tilde\d B\du{\n\r } I$~ in (3.4a).  This sort of sophisticated  
Chern-Simons terms at order $~m^{-1}$~ has not been presented  
in the past, to our knowledge.  This is the result of intricate play 
between the TM and EVM, where the latter absorbs the former as a compensator 
multiplet.  

The confirmation of the action invariance $~\d_Q I = 0$~ is performed as follows. 
We have to confirm this to $~\order\phi2$~ and $~\order\phi3\-$terms, 
where $~\phi$~ stands for any fundamental field. 
To be more precise, there are four categories of terms to consider: 
(I) $m^0 \phi^2$~,~
(II) $m^1 \phi^2$~,~ (III) $m^0 \phi^3$~, and (IV) $m^1 \phi^3$, because of 
the the constant $~m$~ involved.  
 
 The categories (I) and (II) are straightforward quadratic-order confirmations. 
 The category (III) for $m^0 \phi^3\-$terms is non-trivial 
 with  nine sectors:  (i) $~\l\calG H $, ~ (ii) $~ \chi F H $, ~ 
 (iii) $~\l H \cal D \varphi $, ~ (iv) $~ \l \calG L $, ~ (v) $~\r F \calG $, ~ 
 (vi) $~\l L \calD \varphi $, ~ 
 (vii) $~ \chi F L $, ~ (viii) $~\r F \calD \varphi $, and (ix) $~\l \Bar\chi D \r $~ or 
 $~\l \Bar\r D \chi $.  The only subtle sector is (ix), where upon 
 partial integrations, we can get rid of derivatives on $~\l$, such that we 
 are left only with $~\l \chi D \r $~ or $~\l  \r D \chi\-$terms.\footnotes{Here 
 we do {\it not} necessary mean the terms of the type  $~(\Bar\e \g \l) (\Bar\chi \g D \r) $~ or $~(\Bar\e \g \l)  (\Bar\r \g D \chi)$, which are reached {\it after} Fierz arrangements.}  After 
Fierz rearrangements, only the structures 
$~(\Bar\e \g \l) (\Bar\chi \g D \r) $~ and $~(\Bar\e \g \l) 
(\Bar\r \g D \chi)$~ remain, all of which cancel amongst themselves.  
The cancellation confirmation 
of these terms are involved, depending on the number of $~\g\-$matrices 
sandwiched by $~\e$~ and $~\l$.  
This is carried out by adding the non-trivial $~\l \r \-$terms in 
$~\d_Q\chi$, $~\l\chi\-$terms in $~\d_Q \r$, and $~\chi\r\-$terms in $~\d_Q \l$.  
For the category (IV) for $m^1 \phi^3\-$terms has four sectors: 
(i) $ m \l \r^2 $, ~(ii) $ m \l^3$, ~ 
(iii) $ m \l \chi^2$, and (iv) $ m \l\r^2$.  The confirmation of all of these 
sectors are relatively easy, consistently with the $~\l \r \-$terms in 
$~\d_Q\chi$, $~\l\chi\-$terms in $~\d_Q \r$, and $~\chi\r\-$terms in $~\d_Q \l$.

\bigskip\bigskip 




\centerline{\bf 4.~~Field Equations } 
\nobreak 

The field equations in our system are highly non-trivial.  This is due to  
the extra Chern-Simon-type terms in various field strengths.  Even the 
simplest field strength $~\calD_\m\varphi^I$~ has an extra term $~m K\du\m I$.
The explicit forms of our field equations are 
$$ \li{ \fracmm{\d\Lag}{\d A\du\m I} 
\eqdot & - D_\n F^{\m\n \,I} + \fracm 12 \fIJK (\Bar\chi{}^J D^\m \r^K)
	+ \fracm 12 \fIJK (\Bar\r{}^J D^\m \chi^K)
	- \fracm 12 m \fIJK (\Bar\l{}^J \g^\m \l^K) \cr 
& + \fracm 12 \fIJK L\du{\n\r} J \calG^{\m\n\r\, K} 
		- \fracm16 \fIJK \calG\du{\n\r\s} J H^{\m\n\r\s\, K}  
			+ \fIJK L^{\m\n\, J} \calD_\n\varphi^K \eqdot 0 ~~, 
&(4.1\rma) \cr 
\fracmm{\d\Lag}{\d B\du{\m\n} I} \eqdot & + \fracm 12 D_\r \calG^{\m\n\r\, I }
			- \fracm 14 m \fIJK (\Bar\l{}^J \g^{\m\n} \chi^K) 
			+ \fracm 14 \fIJK F\du{\r\s} J H^{\m\n\r\s\, K} \cr 
& - \fracm 12 \fIJK (\Bar\l{}^J \g^{\[\m} D^{\n\]} \r^K) 
		+ \fracm 12 \fIJK (\Bar\r{}^J \g^{\[\m} D^{\n\]} \l^K) \eqdot 0 ~~,  			
&(4.1\rmb) \cr 
\fracmm{\d\Lag}{\d C\du{\m\n\r} I} \eqdot & - \fracm 1 6 D_\s H^{\m\n\r\s\, I}
		- \fracm 16 m \calG^{\m\n\r\, I} 
		- \fracm 16 m \fIJK (\Bar\l{}^J \g^{\m\n\r} \r^K) \cr 
& -  \fracm 14 \fIJK (\Bar\l{}^J \g^{\[\m\n} D^{\r \]} \chi^K )
		+ \fracm 14 \fIJK (\Bar\chi{}^J \g^{\[\m\n} D^{\r \]} \l^K )\eqdot 0 ~~, 
&(4.1\rmc) \cr 
\fracmm{\d\Lag}{\d K\du\m I} \eqdot & - D_\n L^{\m\n\, I} 
- m \calD^\m \varphi^I - \fracm 12 \fIJK F\du{\n\r} J \calG^{\m\n\r\, K} \cr 
& - m \fIJK (\Bar\l{}^J \g^\m \r^K) 
		- \fracm 12\fIJK (\Bar\l{}^J D^\m \chi^K )
		- \fracm 12\fIJK (\Bar\chi{}^J D^\m \l^K ) \eqdot 0 ~~, 
&(4.1\rmd) \cr 
\fracmm{\d\Lag}{\d \varphi^I} \eqdot & + D_\m \calD^\m \varphi^I 
			- \fracm 12 m \fIJK (\Bar\l{}^J \chi^K) 
					+ \fracm 12 \fIJK F\du{\m\n} J L^{\m\n\, K} \eqdot 0 ~~, 
&(4.1\rme) \cr 
\fracmm{\d\Lag}{\d\Bar\l^I} \eqdot & + \Dsl \l^I 
		+ \fracm 1{48} \fIJK (\g^{\m\n\r\s} \chi^J) H\du{ \m\n\r\s } K 
		- \fracm 1{12} \fIJK (\g^{\m\n\r} \r^J) \, \calG\du{\m\n\r} K \cr 
& - \fracm 12 (\g^\m\r^J) \calD_\m \varphi^K 
		+ \fracm 14 (\g^{\m\n} \chi^J) L\du{\m\n} K \eqdot 0 ~~, 
&(4.1\rmf) \cr 
\fracmm{\d\Lag}{\d\Bar\chi^I} \eqdot & + \Dsl \chi^I + m\r^I 
		- \fracm1{48} \fIJK (\g^{\m\n\r\s} \l^J) H\du{\m\n\r\s} K \cr 
& + \fracm 14 \fIJK (\g^{\m\n} \r^J) F\du{\m\n} K 
		+ \fracm 14 \fIJK (\g^{\m\n} \l^J) L\du{\m\n} K \eqdot 0 ~~, 
&(4.1\rmg) \cr 
\fracmm{\d\Lag}{\d\Bar\r^I} \eqdot & + \Dsl \r^I + m \chi^I 
			+ \fracm 1{12} \fIJK (\g^{\m\n\r } \l^J) \, \calG\du{\m\n\r} K \cr 
& - \fracm 14 \fIJK (\g^{\m\n} \chi^J) F\du{\m\n} K 
		- \fracm 14 \fIJK (\g^\m \l^J) \, \calD_\m\varphi^K \eqdot 0 ~~,  
&(4.1\rmh) \cr} $$ 
where the symbol $~\eqdot $~ is for an equality by the use of 
field equation(s).  Also, these equations are valid up to $~\order\phi 3\-$terms.  

The $~m\, \calG\-$term in the $~C\-$field equation (4.1c)  
plays the role of the mass term for the $~C\-$field after a field redefinition 
of $~C$~ absorbing the $~3 D B\-$term in $~\calG$.  
So does the $~m \, \calD\varphi\-$term in the $~K\-$field equation (4.1d).  

The result (4.1) is based on an important lemma about the general variation 
of our lagrangian up to a total divergence: 
$$ \li{ \d \Lag = & (\d A\du\m I) 
	\Bigg[ + 2 D_\n \left( \fracmm{\d\Lag}{\d F\du{\m\n} I} \right) 			
	+ \left( \fracmm{\d\Lag_{\psi\Dsl \psi}}{\d A\du\m I} \right)   
	- 3 \fIJK L\du{\n\r} J \left( \fracmm{\d\Lag}{\d\calG\du{\m\n\r} K} \right) \cr 
& {\hskip 0.8in}  + 4 \fIJK \calG\du{\n\r\s} J 
		\left( \fracmm{\d\Lag}{\d H\du{\m\n\r\s} K } \right)
	+ 2 \fIJK (\calD_\n \varphi^J) \left( \fracmm{\d\Lag}{\d L\du{\m\n} K} \right) 
	\Bigg] \cr 
& + (\d B\du{\m\n} I ) 
	\Bigg[ - 3 D_\r \left(\fracmm{\d\Lag}{\d \calG\du{\m\n\r} I } \right)
			- 6 \fIJK F\du{\r\s} J 
			\left(\fracmm{\d\Lag}{\d H\du{\m\n\r\s} K }\right)  \Bigg] \cr 
& + (\d C\du{\m\n\r} I ) 
	\Bigg[ + 4 D_\s \left(\fracmm{\d\Lag}{\d H\du{\m\n\r\s} I } \right) 
	+ m \left(\fracmm{\d\Lag}{\d\calG\du{\m\n\r} I } \right) \Bigg] \cr 
& + (\d K\du\m I ) 
	\Bigg[ + 2 D_\n \left(\fracmm{\d\Lag}{\d L\du{\m\n} I } \right)
	+ m \left\{ \fracmm{\d\Lag}{\d (\calD_\m\varphi^I ) } \right\} 
	+ 3 \fIJK F\du{\n\r} J 
	\left(\fracmm{\d\Lag}{\d\calG\du{\m\n\r} K } \right) \Bigg] \cr 
& + (\d \varphi^I)
 \Bigg[ - D_\m \left\{ \fracmm{\d\Lag}{\d (\calD_\m\varphi^I) } \right\}  
 		+ 3 m^{-1} \fIJK F\du{\m\n} J 
		D_\r \left(\fracmm{\d\Lag}{\d\calG\du{\m\n\r} K} \right)
	- \fIJK F\du{\m\n} J \left(\fracmm{\d\Lag}{\d L\du{\m\n} K} \right) \Bigg] \cr 
& + (\d\Bar\l{}^I ) \left( \fracmm{\d\Lag}{\d\Bar\l{}^I} \right) 
+ (\d\Bar\chi{}^I ) \left( \fracmm{\d\Lag}{\d\Bar\chi{}^I} \right) 
+ (\d\Bar\r{}^I ) \left( \fracmm{\d\Lag}{\d\Bar\r{}^I} \right) 
&(4.2) \cr } $$  	 
The symbol $~( {\d\Lag_{\psi\Dsl\psi}}/ {\d A\du\m I} )$~ in the first line is for 
the contributions from the minimal couplings in the fermionic kinetic terms of 
$~\l,~\chi$~ and $~\r$. 
Use is also made of the general-variation formulae in (3.3) for arranging the
whole terms.  

In getting the expression (4.2), there are many non-trivial cancellations.  
For example, the two terms:  
$$ \li{ & 3 \fIJK (\d A\du\m I )B\du{\n\r} J  \left[ \, 
+ 4 D_\s \left(\fracmm{\d\Lag}{\d H\du{\m\n\r\s} K } \right)
+ m \left( \fracmm{\d\Lag}{\d\calG\du{\m\n\r}K} \right) \, \right] 
&(4.3) \cr} $$  
cancel up to $~\order\phi3$~ 
upon the use of the $~C\-$field equation (4.1c).  
Similarly, the two terms:  
$$ \li{ & \fIJK (\d A\du\m I )\, \varphi^J  \left[\, 
+ 2 D_\n \left( \fracmm{\d\Lag}{\d L\du{\m\n} K } \right)
+ m \left\{ \fracmm{\d\Lag}{\d(\calD_\m\varphi^K)} \right \} \, \right] 
&(4.4) \cr} $$  
also cancel upon the $~K\-$field equation (4.1d) up to $~\order\phi3$.   

By straightforward computations, we can confirm that the supersymmetric 
variation of each of the field equations in (4.1) vanishes up to $~\order\phi 3$.  
This gives an independent confirmation of the consistency of our total system.   

As an additional confirmation, we can show that the divergence of 
the $~A,~B,~C$~ and $~K\-$field equations all vanish.  
For example, the divergence of the $~A\-$field equation is 
$$ \li{ 0 \eqques & \! D_\m \left( \fracmm{\d\Lag}{\d A\du\m I} \right) 
	\eqdot + m \fIJK (\Bar\chi{}^J \r^K) + m \fIJK (\Bar\r{}^J \chi^K) 
	- \fracm1{24} m \fIJK H\du{\m\n\r\s} J H^{\m\n\r\s\, K}  \cr  
&~~~~~    - \fracm 16 \fIJK \calG\du{\n\r\s} J \calG^{\n\r\s\, K}
		- \fracm 12 m \fIJK L\du{\m\n} J L^{\m\n\, K} + \order\phi 3 
		\eqdot \order\phi3 ~~~~ \qed~~. ~~~~~ ~~~~~ ~~ 
&(4.5) \cr} $$ 
Here we have used other field equations, such as $~\Dsl \l^I \eqdot \order\phi2$~
or $~D_\m H^{\m\n\r\s\, I} \eqdot + m \calG^{\n\r\s\, I}+ \order\phi2$, {\it etc.}  
Similarly for the case of $~C\-$field equation:
$$ \li{ 0 \eqques \! & D_\r \left( \fracmm{\d\Lag}{\d C\du{\m\n\r} I} \right) 
		= - \fracm 1{12} m \fIJK F\du{\r\s} J H^{\m\n\r\s\, K} 
	- \fracm 16 m D_\r \calG^{\m\n\r} 
	- \fracm1{12} m \fIJK D_\r (\Bar\l{}^J \g^{\m\n\r} \r^K) + \order\phi 3 \cr 
\eqdot & - \fracm 1{12} \fIJK F\du{\r\s} J H^{\m\n\r\s \, K} 
		- \fracm 16 m \left[ \, - \fracm 12 \fIJK D_\r ( \Bar\l{}^J \g^{\m\n\r} \r^K )
		- \fracm 12 \fIJK F\du{\r\s} J H^{\m\n\r\s\, K} \, \right] \cr 
& - \fracm 1{12} m \fIJK D_\r (\Bar\l{}^J \g^{\m\n\r} \r^K) + \order\phi3
		\eqdot \order\phi3~~,  
&(4.6) \cr} $$ 
where we used the $~B\-$field equation for the $~D \, \calG\-$term.


\bigskip\bigskip



\centerline{\bf 5.~~Parity-Odd Terms} 
\nobreak 

We can add certain parity-odd terms to our original lagrangian $~\Lag$.
The total invariant action is $I_{\rm tot} \equiv I + I_{\a,\,\b}$~  
with $~I_{\a,\,\b} \equiv \int d^4 x\, \Lag_{\a,\,\b}$~ with  
arbitrary real constants $~\a$~ and $~\b$:  
$$ \li{ \Lag_{\a,\,\b} \equiv & + \fracm 18 \a\, \e^{\m\n\r\s} L\du{\m\n} I L\du{\r\s} I 
		+ \fracm 16 \b \, \e^{\m\n\r\s} \calG\du{\m\n\r} I \calD_\s \varphi^I \cr 
&  + \fracm 1 4(\a+\b) \, \e^{\m\n\r\s} \fIJK F\du{\m\n} I
			 (\calD_\r\varphi^J) (\calD_\s\varphi^K) 
			 - i m \b \, (\Bar\chi{}^I \g_5 \r^I) ~~.  
&(5.1) \cr} $$  
The $~\order\phi2\-$term of the $~\a L\wedge L\-$term is a total divergence.  
The $~\a\, \calG \wedge \calD \,\varphi\-$term has both $~\order\phi2$~ 
and $~\order\phi3\-$terms, the former of which cancels the like terms from $~ i m\a 
(\Bar\chi^I\g_5 \r^I)$~ under the variation $~\d_Q$.  
The original transformation rule (3.2) is {\it not} modified by $~\a$~ or $~\b$.  
The $~\a L \wedge L\-$term 
is an analog of the $~\theta F \wedge F\-$term associated with the 
$~\theta\-$vacuum in QCD   
\ref\thetavacuum{\thetavacuumcont},   
or the $~U(1)_{\rm A}$~ problem 
\ref\uone{\uonecont}.  
However, our $~\a L \wedge L\-$term is more involved, 
because of the non-trivial Chern-Simons term $~F \varphi$~ in $~L$.  

The invariance $~\d_Q I_{\rm tot} = 0 $, and in particular $~\d_Q I_{\a,\,\b}=0$~  
up to $~\order\phi 4$~ is easily confirmed.  Even the non-trivial looking (fermions)$^3\-$terms in the variation turn out to be simple, because of the algebra,  
$~\fIJK (\Bar\chi{}^J \g_5 \g^\m \chi^K ) = \fIJK (\Bar\chi{}^J \chi^K ) 
= \fIJK (\Bar\chi{}^J \g_5 \chi^K ) \equiv 0$, {\it etc.}  
  
\bigskip\bigskip



\centerline{\bf 6.~~Concluding Remarks} 
\nobreak 

In this paper, we have presented a very peculiar supersymmetric system 
that realizes the Proca-Stueckelberg compensator mechanism \stueckelberg\ 
for an extra vector multiplet.    
Our present model has resemblance to our recent model \nrstueckel, 
which had only two multiplets VM and TM.  

The peculiar features of our model are summarized as 

\vskip -0.03in 
\noindent 
{\bf (i)}~~We have three multiplets VM, TM and EVM, where the EVM will 
be eventually massive.  

\vskip -0.03in 
\noindent
{\bf (ii)}~~Our peculiar field strength $~\calG = 3 D B 
+ m C - 3m^{-1} F \wedge B$~ has the last term with $~m^{-1}$.  

\vskip -0.03in 
\noindent 
{\bf (iii)}~~Our model provides yet another mechanism of absorbing the 
dilaton-type scalar field $~\varphi^I$~ into the extra vector $~K\du\m I$, 
different from the conventional YM gauge field $~A\du\m I$.  

\vskip -0.03in 
\noindent
{\bf (iv)}~~Even the tensor $~C\du{\m\n\r}I$~ in the EVM 
gets a mass absorbing $~B\du{\m\n}I$~ in the TM.   

\vskip -0.03in 
\noindent 
{\bf (v)}~~Our system accommodates also parity-odd terms, analogous to the 
$~\theta F\wedge F\-$term \thetavacuum\uone.

Even though our system is less economical than \nrstueckel\ with an additional 
multiplet EVM, it has its own advantage.  First, we provide a mechanism for giving a mass to the extra vector $~K\du\m I$~ in the EVM, 
which may be not needed as a massless particle at low energy.  Second,  
we have a new compensator mechanism for an extra vector in the adjoint representation, which is {\it not} the YM gauge field.  The derivative 
$~\calD_\m\varphi^I$~ is simpler than exponentiations \sw\nrstueckel. 


General formulations for different representations 
(not necessarily adjoint representations) for 
supersymmetric compensator mechanism have been given 
in \sw.  However, we emphasize here that 
the fixing of supersymmetric couplings for 
our system with a different field content is a highly non-trivial task.   
Even superspace formulation does not help so much, 
as described in \nrstueckel.  The main reason is that 
the usual unconstrained formalism in terms of the singlet superfield $~L$~ 
\ref\muller{\mullercont}
can {\it not} describe a tensor multiplet in the adjoint representation.  

Our results can be applied to diverse dimensions and also to extended supersymmetric systems.


\doit1{\bigskip 
This work is supported in part by Department of Energy 
grant \# DE-FG02-10ER41693.  
} 

\bigskip\bigskip\bigskip 



\def\texttts#1{\small\texttt{#1}}

\immediate\closeout\rfile\writestoppt
\baselineskip=12.5pt\centerline{{\bf References}}
\font\smallerfonts=cmr10 \font\it=cmti10 \font\bf=cmbx10%
\bigskip{\smallerfonts{%
\parindent=18pt\escapechar=` \input refs.tmp\vfill\eject}}


\vfill\eject

\end{document}